\documentclass[aps,prd,twocolumn,showpacs,nofootinbib,nonfootnoteinbib]{revtex4-1}

\usepackage{graphicx}

\begin{document}

\title{Hadronic production of $\Xi_{cc}$ at a fixed-target experiment at the LHC}

\author{Gu Chen$^1$}
\author{Xing-Gang Wu$^1$}
\email{email:wuxg@cqu.edu.cn}
\author{Jia-Wei Zhang$^2$}
\author{Hua-Yong Han$^1$}
\author{Hai-Bing Fu$^1$}

\address{$^1$Department of Physics, Chongqing University, Chongqing 401331, P.R. China.\\
$^2$Department of physics, Chongqing University of Science and Technology, Chongqing 401331, P.R. China.}

\date{\today}

\begin{abstract}

In the paper, we present a detailed discussion on the $\Xi_{cc}$ production at a fixed target experiment at the LHC (After@LHC). The doubly charmed baryon $\Xi_{cc}$ is produced via the channel, ${\rm Proton} + {\rm Proton}\to\Xi_{cc}+X$. In estimating its hadroproduction, we discuss three dominant subprocesses, e.g. $g+g\to \Xi_{cc} +\bar{c} +\bar{c}$, $g+c\to \Xi_{cc}+\bar{c}$ and $c+c\to \Xi_{cc}+g$. During the production, it shall first generate a binding diquark and then form the $\Xi_{cc}$ baryon by grabbing soft light-quarks or gluons. We observe that both the two diquark configurations $(cc)[^3S_1]_{\bf\bar 3}$ and $(cc)[^1S_0]_{\bf 6}$ can have sizable contributions to the $\Xi_{cc}$ production. Large number of $\Xi_{cc}$ events can be generated at the After@LHC, whose total production cross section is larger than that of the SELEX experiment by about thirty-five times. It may also possible to study the properties of $\Xi_{bc}$ at the After@LHC. More specifically, we shall have about $8.3 \times 10^6$ $\Xi_{cc}$ events/year and $1.8 \times 10^4$ $\Xi_{bc}$ events/year when its integrated luminosity approaches to $2$ fb$^{-1}$/year. Thus, in addition to SELEX and LHC, the After@LHC shall provide another useful platform for studying the baryon properties. \\


\end{abstract}

\maketitle

\section{Introduction}

Since its discovery by the SELEX collaboration~\cite{selex1,selex2}, the doubly charmed baryon $\Xi^+_{cc}$ has attracted more and more attentions. A recent experimental research for $\Xi_{cc}$ has been done by the LHCb collaboration at the LHC~\cite{lhcb}. The SELEX measurements indicate that there are large discrepancies between the theoretical prediction and the experimental observation. Its measured decay widths and production rates are much larger than the theoretical predictions. Lots of theoretical works have been tried to resolve such discrepancy~\cite{the1,the2,baranov,kiselev1,kiselev2,kiselev3,saleev,genxicc,xicc2,xicc6}. Especially, a generator GENXICC has been programmed to simulate the production of the doubly heavy baryons at the hadronic colliders as Tevatron and LHC~\cite{xicc3,xicc4,xicc5}. By using GENXICC, it has been stated that $\Xi_{cc}$ should be visible if an integrated luminosity $5$ fb$^{-1}$ has been collected at the LHC~\cite{lhcbexp}. However, due to the present limited experimental measurements, the puzzle is still there.

On the one hand, we need to improve our present theoretical estimations by studying all possible quark configurations for constructing the baryon and by taking more production or decay channels into consideration. On the other hand, in addition to SELEX, it is helpful to find some other platforms which can generate large number of baryons so as to study their properties with higher precision. For example, the possibilities for studying the doubly heavy baryons at a high luminosity $e^+ e^-$ collider or a photon-photon collider have been suggested in Refs.\cite{zxicc,zxicc1,majp,zxicc2,zxicc3}. A high luminosity $e^+ e^-$ collider or a photon-photon collider has some advantages for measuring the baryon events, such as the cleanness of the physical background and etc.. Thus, if sizable baryon events can be produced at such colliders, they shall provide a good platform for testing the QCD factorization theory.

For the baryon production at the hadronic platforms, the baryons can be produced through scattering, annihilating, or fusing of two initial partons inside the incident hadrons. In comparison to the $e^+ e^-$ collider, the hadronic production are complicated due to the entanglement of the perturbative kernel with the nonperturbative parton distribution functions (PDFs). But, inversely, if we have known more accurate baryon properties in comparison with the experimental data, we can learn more details on the properties of hadron structures. So, to study the baryon productions at the hadronic colliders are also helpful and interesting.

Recently, similar to the SELEX experiment done at the Tevatron, another fixed target experiment at the LHC (After@LHC) has been suggested~\cite{after1,after2,after3,after4}. With the incident proton beam energy raises up to $7$ TeV at the LHC, the After@LHC shall run with the center-of-mass energy around $115$ GeV. With a much higher luminosity and higher collision energy, the After@LHC shall become a much better fixed-target experiment for studying the properties of the doubly heavy baryons. In the present paper, we shall present a detailed investigation on the $\Xi_{cc}$ production at the After@LHC by taking both the gluon-gluon fusion mechanism via the subprocess $g+g\to \Xi_{cc} +\bar{c} +\bar{c}$ and the the extrinsic charm mechanism via the subprocesses $g+c\to \Xi_{cc}+\bar{c}$ and $c+c\to \Xi_{cc}+g$ into consideration. As a by product, we shall also estimate the production properties for the $\Xi_{bc}$ and $\Xi_{bb}$ baryons.

The remaining parts of the paper are organized as follows. In Sec.II, we explain our calculation technology for dealing with the $\Xi_{cc}$ production. In Sec.III, we present our numerical results for the doubly heavy baryons $\Xi_{cc}$, $\Xi_{bc}$ and $\Xi_{bb}$. Sec.IV is reserved for a summary.

\section{Hadronic production of $\Xi_{cc}$}

Theoretically, the production of the doubly heavy baryons, such as $\Xi_{cc}$, $\Xi_{bc}$, and $\Xi_{bb}$, can be treated within the framework of nonrelativistic QCD (NRQCD)~\cite{nrqcd}.

It is usually assumed that the doubly heavy baryon can be formed via a step-by-step way. The first step is to produce two free heavy-quark pairs $Q\bar{Q}$ and $Q^{\prime}\bar{Q}^{\prime}$, where $Q$ or $Q^{\prime}$ stands for $b$ or $c$ quark, respectively. The gluon between those two heavy-quark pairs should be hard enough to generate either a $Q\bar{Q}$-pair or a $Q^{\prime}\bar{Q}^{\prime}$-pair, so this step is calculable by applying pQCD. The second step is to make those two heavy quarks $Q$ and $Q^{\prime}$ into a bounding diquark $(QQ^{\prime})$ with the $[^3S_1]$ (or $[^1S_0]$) spin state and the $\mathbf{\bar{3}}$ (or $\mathbf{6}$) color state, respectively. Then, the diquark will be hadronized into a $\Xi_{QQ^{\prime}}$ baryon, whose probability is described by the non-perturbative NRQCD matrix element $h_1$ or $h_3$. Here, $h_1$ represents the probability for a $(QQ')$-diquark pair in $(QQ')_{\bf 6}[^1S_0]$ to transform into the baryon and $h_3=|\Psi_{QQ'}(0)|^2$ represents the probability for a $(QQ')$-diquark pair in $(QQ')_{\bf\bar{3}}[^3S_1]$ to transform into the baryon. Qualitatively, the values of $h_1$ and $h_3$ are at the same order~\cite{majp}, so we shall take $h_1=h_3$ to do our following discussion. For $\Xi_{bc}$, there are other two diquark states, $(bc)_{\bf 6}[^3S_1]$ and $(bc)_{\bf\bar{3}}[^1S_0]$, we shall also take their matrix elements as $|\Psi_{bc}(0)|^2$. Strictly, during the fragmentation of a diquark into a baryon, the diquark may dissociate, which will decrease the baryon production cross section to a certain degree. It has been observed that the fragmentation function $D(z)$ of a heavy diquark into a baryon peaks around $z\approx 1$~\cite{kiselev1}. By taking a simple form for the fragmentation function $D(z)$, Refs.\cite{kiselev2,saleev} indicate that such disassociation effect is small. Thus, at present, we have implicitly assumed that the fragmentation of a diquark into the baryon shall occur with unit probability, and consequently, to study the hadronic production of $\Xi_{cc}$ is equivalent to study the hadronic production of $(cc)$-diquark. A detailed discussion on such disassociation effect is in preparation. Our present estimations can be treated as a (somewhat good) upper limit for the total baryon cross sections.

\begin{figure}[tb]
\includegraphics[width=0.45\textwidth]{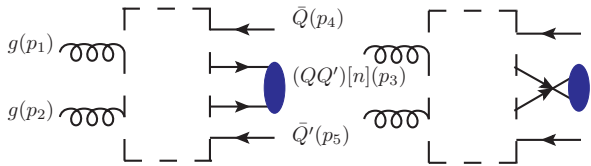}
\caption{The schematic Feynman diagrams for the baryon production via the gluon-gluon fusion mechanism, $g(p_{1})+g(p_{2}) \to \Xi_{QQ'}(p_3) + \bar{Q}(p_{4})+\bar{Q'}(p_{5})$ with the diquark state $(QQ')[n]$, where $Q$ or $Q^{\prime}$ stands for $b$ or $c$ quark, respectively. The dashed boxes stand for the hard interaction kernel, each contains $36$ Feynman diagrams.} \label{gg}
\end{figure}

\begin{figure*}
\includegraphics[width=0.9\textwidth]{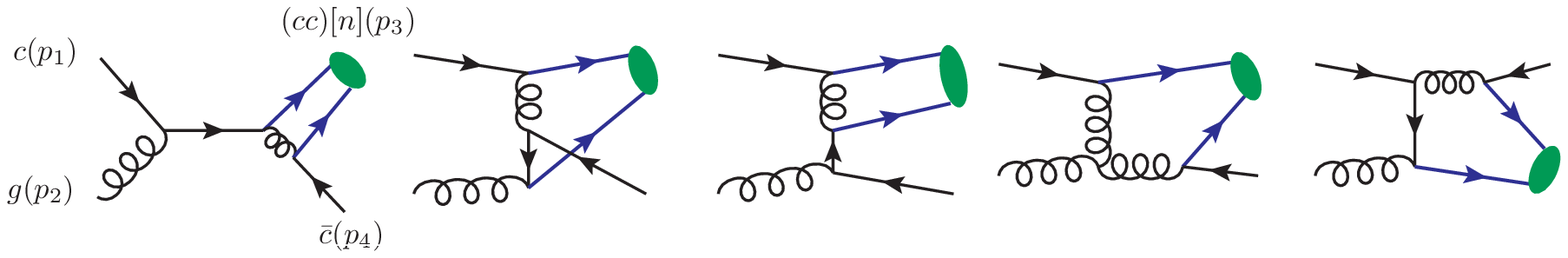}
\caption{Typical Feynman diagrams for the process $c(p_{1})+g(p_{2}) \to \Xi_{cc}(p_{3})+ \bar{c}(p_{4})$ via the diquark state $(cc)[n]$, where the intermediate $(cc)$-diquark is in $[^3S_1]_{\bar{\textbf{3}}}$ or $[^1S_0]_{\textbf{6}}$, respectively. } \label{gc}
\end{figure*}

\begin{figure*}
\includegraphics[width=0.9\textwidth]{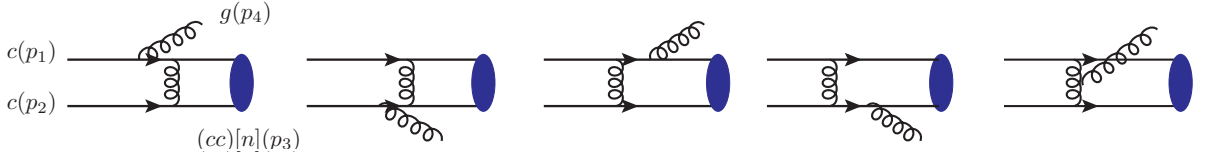}
\caption{Typical Feynman diagrams for the process $c(p_{1})+c(p_{2}) \to \Xi_{cc}(p_{3})+ \bar{c}(p_{4})$ via the diquark state $(cc)[n]$, where the intermediate $(cc)$-diquark is in $[^3S_1]_{\bar{\textbf{3}}}$ or $[^1S_0]_{\textbf{6}}$, respectively.} \label{cc}
\end{figure*}

There are two typical mechanisms for the production of baryons, i.e. the gluon-gluon fusion mechanism via the subprocess $g+g\to \Xi_{cc} +\bar{c} +\bar{c}$ and the the extrinsic charm mechanism via the subprocesses $g+c\to \Xi_{cc}+\bar{c}$ and $c+c\to \Xi_{cc}+g$. The subprocesses for the production of doubly heavy baryon through different mechanisms are shown in Figs. \ref{gg}, \ref{gc}, and \ref{cc}, respectively. By simultaneously taking the gluon-gluon fusion and extrinsic charm mechanisms into consideration, one will meet the double counting problem~\cite{double1,double2}, which can be treated within the general-mass with variable-flavor-number scheme~\cite{gmvfn1,gmvfn2,gmvfn3}.

The hadronic production conditions for the $\Xi_{bc}$ and $\Xi_{bb}$ baryons are similar to the $\Xi_{cc}$ case, so we shall also take a look at their production properties. For $\Xi_{bc}$ and $\Xi_{bb}$ production, we merely focus on the gluon-gluon fusion mechanism, since the extrinsic mechanisms of them shall provide comparatively small contributions. It is noted that the intermediate diquarks in $\Xi_{cc}$ and $\Xi_{bb}$ have two spin and color configurations, $[^3S_1]_{\bf\bar{3}}$ and $[^1S_0]_{\bf 6}$; while for the intermediate diquark $(bc)$ in $\Xi_{bc}$, there are four spin and color configurations: $\Xi_{bc}[^3S_1]_{\bf\bar{3}}$, $\Xi_{bc}[^3S_1]_{\bf 6}$, $\Xi_{bc}[^1S_0]_{\bf\bar{3}}$, and $\Xi_{bc}[^1S_0]_{\bf 6}$. We shall take all those configurations into consideration, since all of them may provide sizable contributions.

\section{Numerical results and discussions}

We adopt GENXICC program~\cite{xicc3,xicc4,xicc5} with slight changes to do our calculation. As for the input parameters, we take~\cite{baranov}: $|\Psi_{cc}(0)|^2 =0.039$ GeV$^3$, $|\Psi_{bc}(0)|^2=0.065$ GeV$^3$, and $|\Psi_{bb}(0)|^2 = 0.152$ GeV$^3$. As for the baryon masses, we take $M_{\Xi_{cc}}=3.50$ GeV with $m_c=M_{\Xi_{cc}}/2$, $M_{\Xi_{bc}}=6.9$ GeV with $m_c=1.8$ GeV and $m_b=5.1$ GeV, $M_{\Xi_{bb}}=10.2$ GeV with $m_b=M_{\Xi_{bb}}/2$, respectively. We choose the CTEQ with the version CT10~\cite{ct10} for the PDF of the gluon or the quarks.

\begin{table}[htb]
\begin{tabular}{|c|c|c|c|}
\hline
~~ ~~ & ~$\sigma_{g+g}$ (pb)~ & ~$\sigma_{g+c}$ (pb)~ & ~$\sigma_{c+c}$ (pb)~ \\
\hline
$(cc)_{\bar{\textbf{3}}}[^3S_1] $ & 530 & 3.19$\times10^3$ & 0.999 \\
\hline
$(cc)_{\textbf{6}}[^1S_0] $  & 99.7 & 348 & 0.040 \\
\hline
\end{tabular}
\caption{Total cross sections for the $\Xi_{cc}$ production at the After@LHC with $\sqrt{S}\simeq 115$ GeV, where the intermediate $(cc)$-diquark is in $[^3S_1]_{\bar{\textbf{3}}}$ or $[^1S_0]_{\textbf{6}}$, respectively. $m_c=1.75$ GeV and $p_{t}>0.2$ GeV. } \label{tablecc}
\end{table}

\begin{table}[htb]
\begin{tabular}{|c|c|c|c|c|}
\hline
~ ~ & $(bc)_{\bar{\textbf{3}}}[^3S_1]$ & $(bc)_{\textbf{6}}[^1S_0]$ & $(bc)_{\textbf{6}}[^3S_1]$ & $(bc)_{\bar{\textbf{3}}}[^1S_0]$ \\
\hline
$\sigma_{g+g}$ (pb) & 2.63 & 0.698 & 4.87 & 0.747 \\
\hline
\end{tabular}
\caption{Total cross sections for the $\Xi_{bc}$ production at the After@LHC with $\sqrt{S}\simeq115$ GeV, where the intermediate ($bc$)-diquark is in $[^3S_1]_{\bar{\textbf{3}}}$, $[^1S_0]_{\textbf{6}}$, $[^3S_1]_{\textbf{6}}$, or $[^1S_0]_{\bar{\textbf{3}}}$, respectively. $m_c=1.80$ GeV, $m_b=5.10$ GeV, and $p_{t}>0.2$ GeV. }
\label{tablebc}
\end{table}

\begin{table}[htb]
\begin{tabular}{|c|c|c|}
\hline
~~ ~~ & ~$(bb)_{\bar{\textbf{3}}}[^3S_1]$~ & ~$(bb)_{\textbf{6}}[^1S_0]$~ \\
\hline
$\sigma_{g+g}$ (pb) & 0.026 & 0.005 \\
\hline
\end{tabular}
\caption{Total cross sections for the $\Xi_{bb}$ production at the After@LHC with $\sqrt{S}\simeq 115$ GeV, where the intermediate $(bb)$-diquark is in $[^3S_1]_{\bar{\textbf{3}}}$ or $[^1S_0]_{\textbf{6}}$, respectively. $m_b=5.10$ GeV and $p_{t} > 0.2$ GeV. }
\label{tablebb}
\end{table}

We put the numerical results for the baryon production in Tables \ref{tablecc}, \ref{tablebc}, and \ref{tablebb}, in which we have taken a small $p_t$ cut as the same as the SELEX experiment, i.e. $p_t > 0.2$ GeV, to do our calculation. The collision energy $\sqrt{S}\simeq 115$ GeV. By summing up all the mentioned states and production channels together, we obtain
\begin{eqnarray}
\sigma_{\rm tot}(\Xi_{cc}) &=& 4.17\times10^{3}\; {\rm pb}, \\
\sigma_{\rm tot}(\Xi_{bc}) &=& 8.95\; {\rm pb}, \\
\sigma_{\rm tot}(\Xi_{bb}) &=& 31\; {\rm fb}.
\end{eqnarray}
Supposing the integrated luminosity at the After@LHC can be reached up to $0.05~{\rm fb}^{-1}$/year or $2~{\rm fb}^{-1}$/year~\cite{after3}, we shall have $2.1\times10^5$ or $8.3\times10^6$ $\Xi_{cc}$ events per year;  $4.5\times10^2$ or $1.8\times10^4$ $\Xi_{bc}$ events per year; $1.6$ or $62$ $\Xi_{bb}$ events per year. This shows that at the After@LHC, sizable $\Xi_{cc}$ and $\Xi_{bc}$ events can be generated, but it is hard to observe $\Xi_{bb}$ events.

\begin{figure}[htb]
\includegraphics[width=0.5\textwidth]{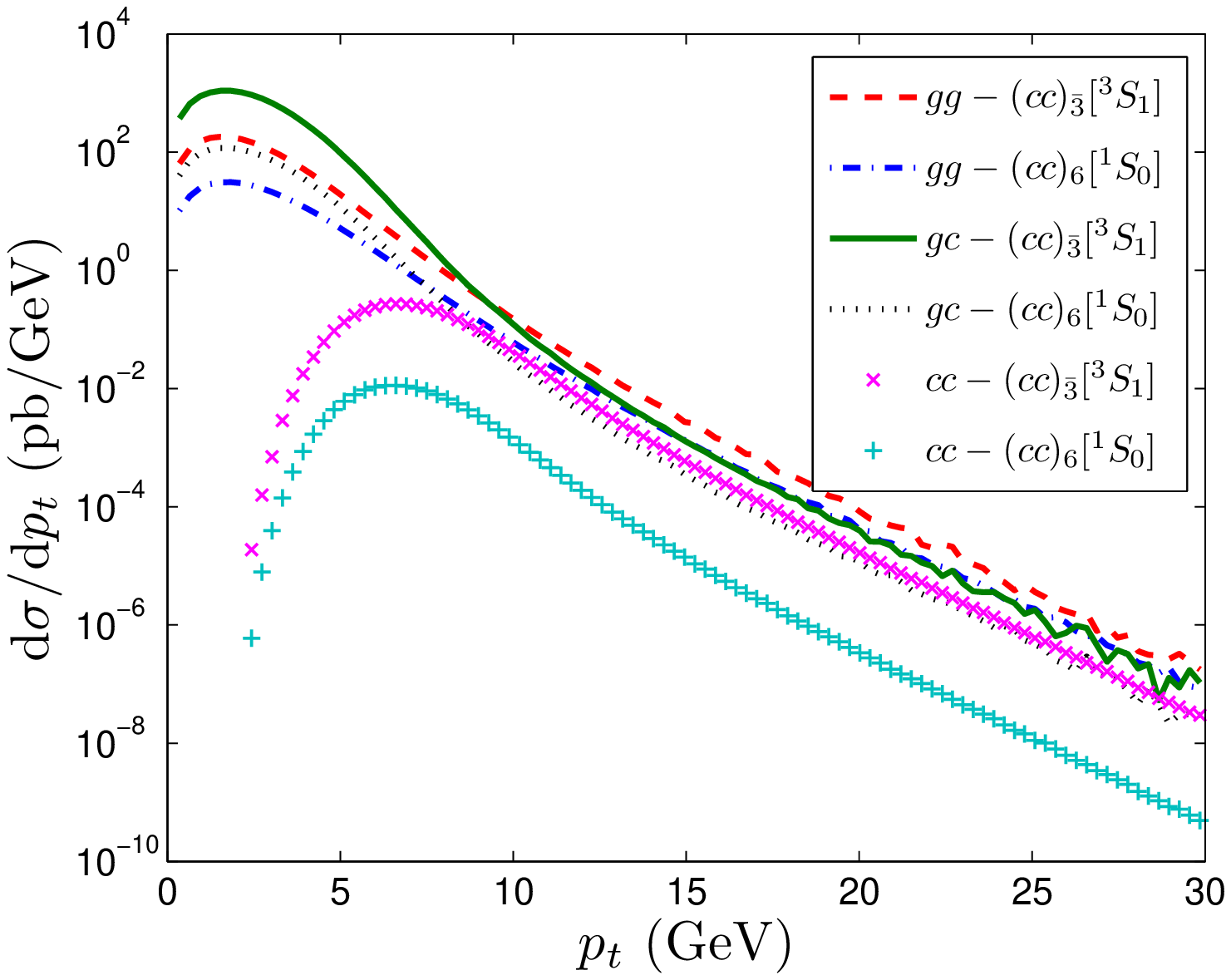}
\includegraphics[width=0.5\textwidth]{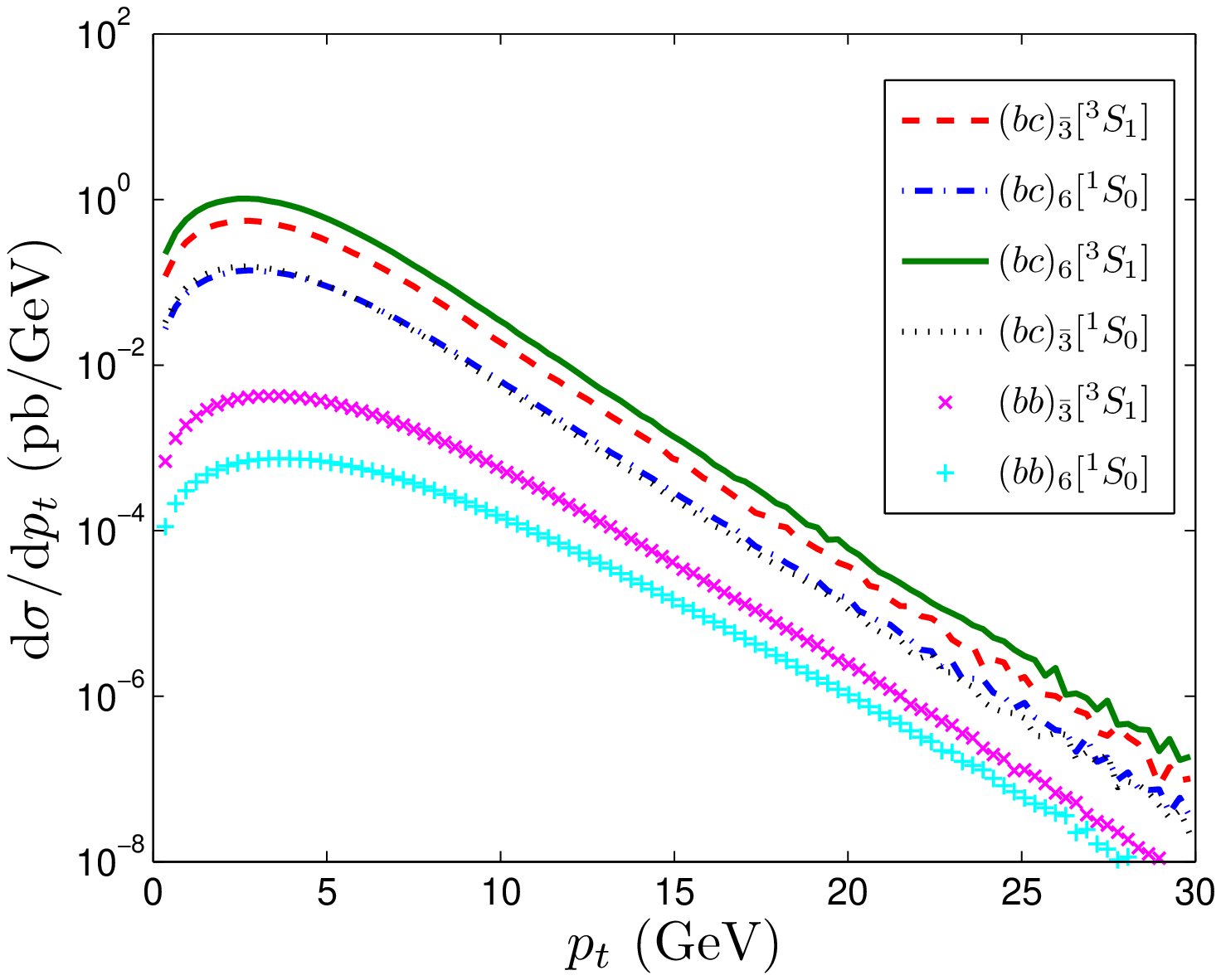}
\caption{The $\Xi_{cc}$, $\Xi_{bc}$, and $\Xi_{bb}$ $p_t$-distributions for various intermediate diquark states at the After@LHC, in which no rapidity cut has been applied. }  \label{ptcc}
\end{figure}

\begin{figure}[htb]
\includegraphics[width=0.5\textwidth]{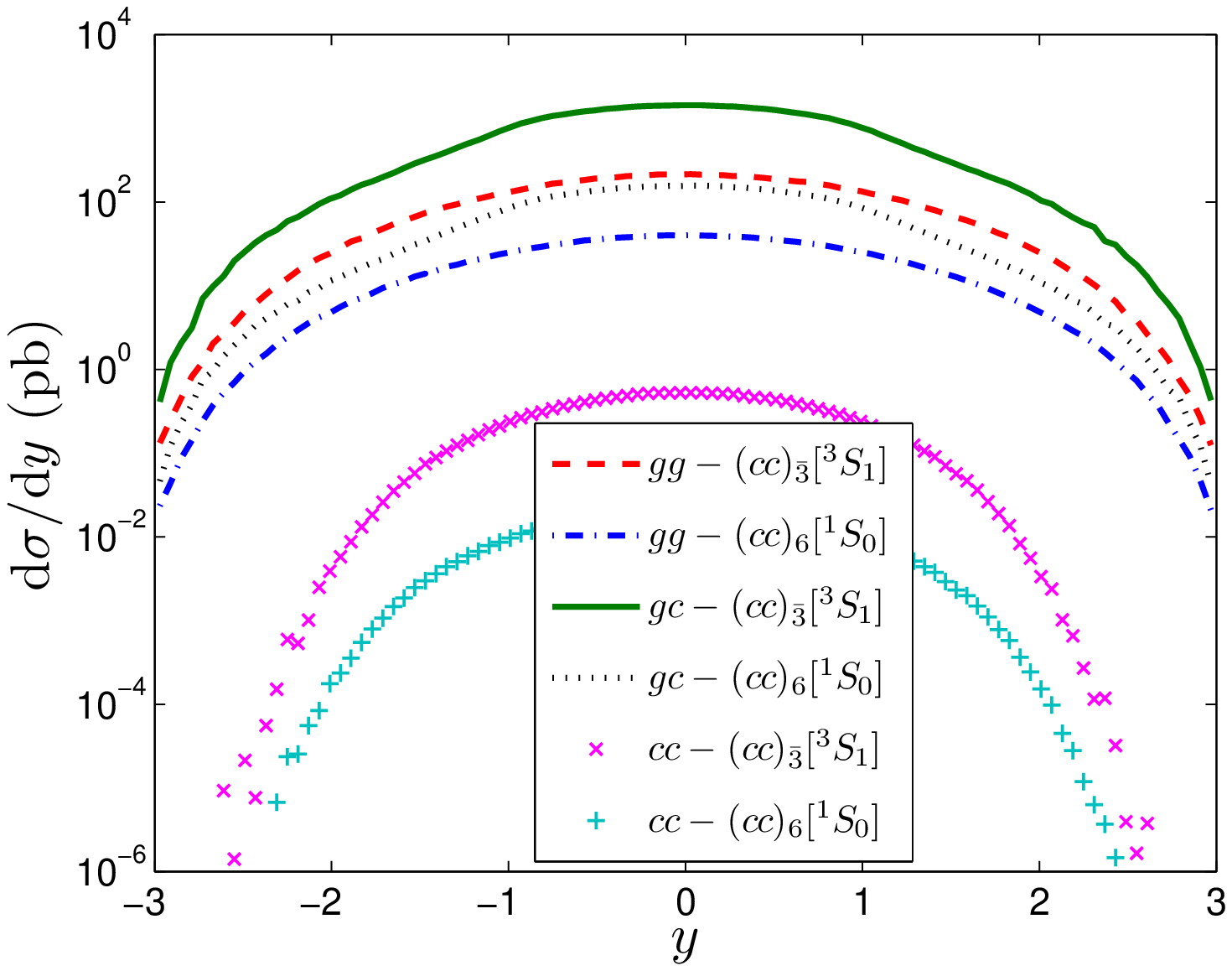}
\includegraphics[width=0.5\textwidth]{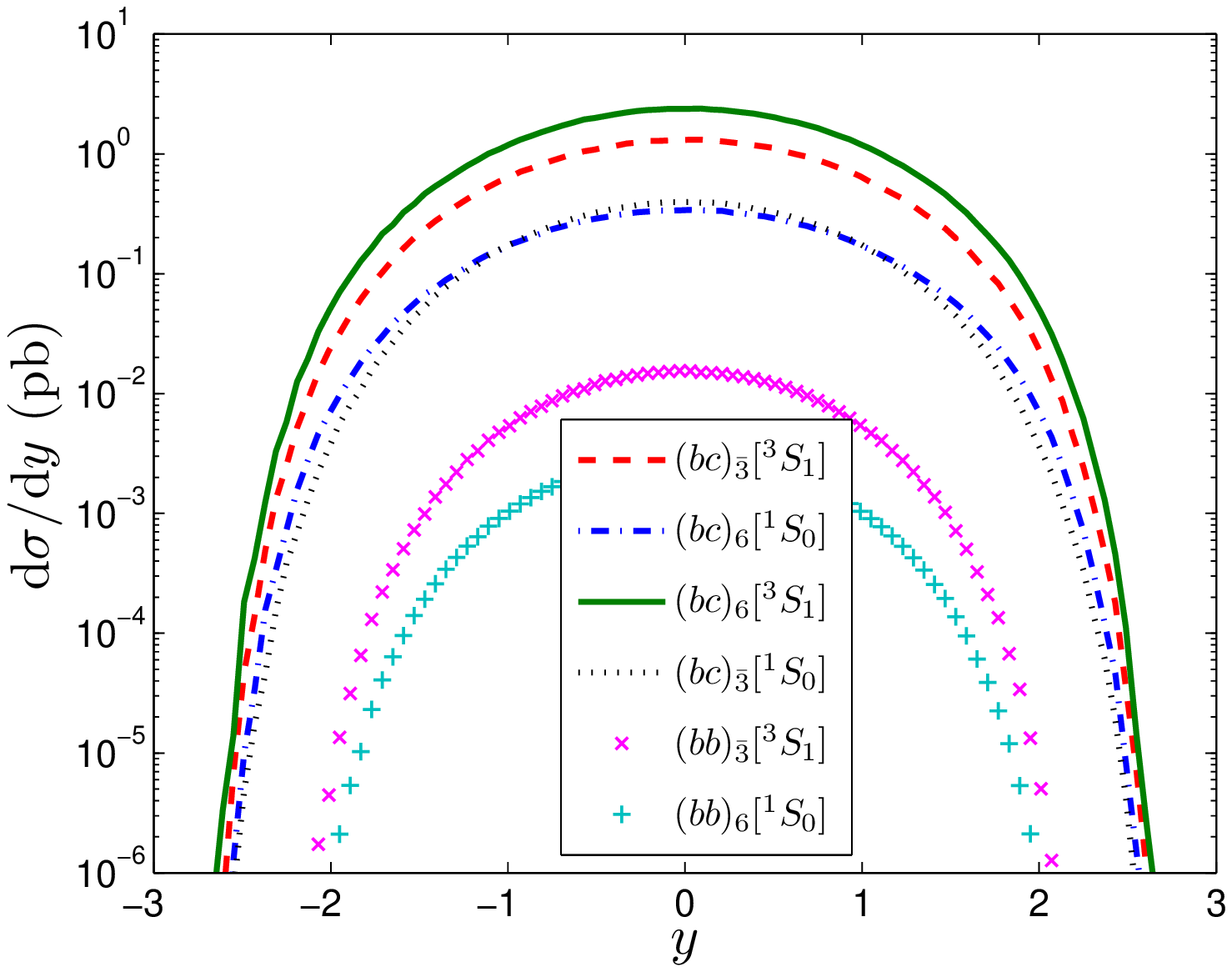}
\caption{The $\Xi_{cc}$, $\Xi_{bc}$, and $\Xi_{bb}$ rapidity distributions for various intermediate diquark states at the After@LHC, where the $p_t>0.2$ GeV is taken. }   \label{ycc}
\end{figure}

We present the baryon transverse momentum ($p_t$)-distributions at the After@LHC in Fig.(\ref{ptcc}) and the baryon rapidity ($y$)-distributions at the After@LHC in Fig.(\ref{ycc}). There are several production channels for the $\Xi_{cc}$ production, i.e., $g+g$, $g+c$, and $c+c$, all of which shall have sizable contributions. When summing up the contributions from different $(cc)$-diquark configurations, one obtains the relative importance among different production channels,
\begin{displaymath}
\sigma_{g+g}: \sigma_{g+c}: \sigma_{c+c}\simeq 6.1\times10^{2} : 3.4\times10^{3} : 1 .
\end{displaymath}
The contribution from $(cc)[^3S_1]_{\bar{\textbf{3}}}$ dominates over that of $(cc)[^1S_0]_{\textbf{6}}$ by about five times for the $g+g$ channel, nine times for the $g+c$ channel, and twenty-five times for the $c+c$ channel, respectively. Similar to the case of SELEX~\cite{genxicc}, the small $p_t$ behavior of the extrinsic charm mechanism becomes very important at the After@LHC. This also indicates the importance of such kind of fixed target experiments, in which it can reach quite small $p_t$ region and know more detail on hadron structures.

\begin{figure}[htb]
\includegraphics[width=0.45\textwidth]{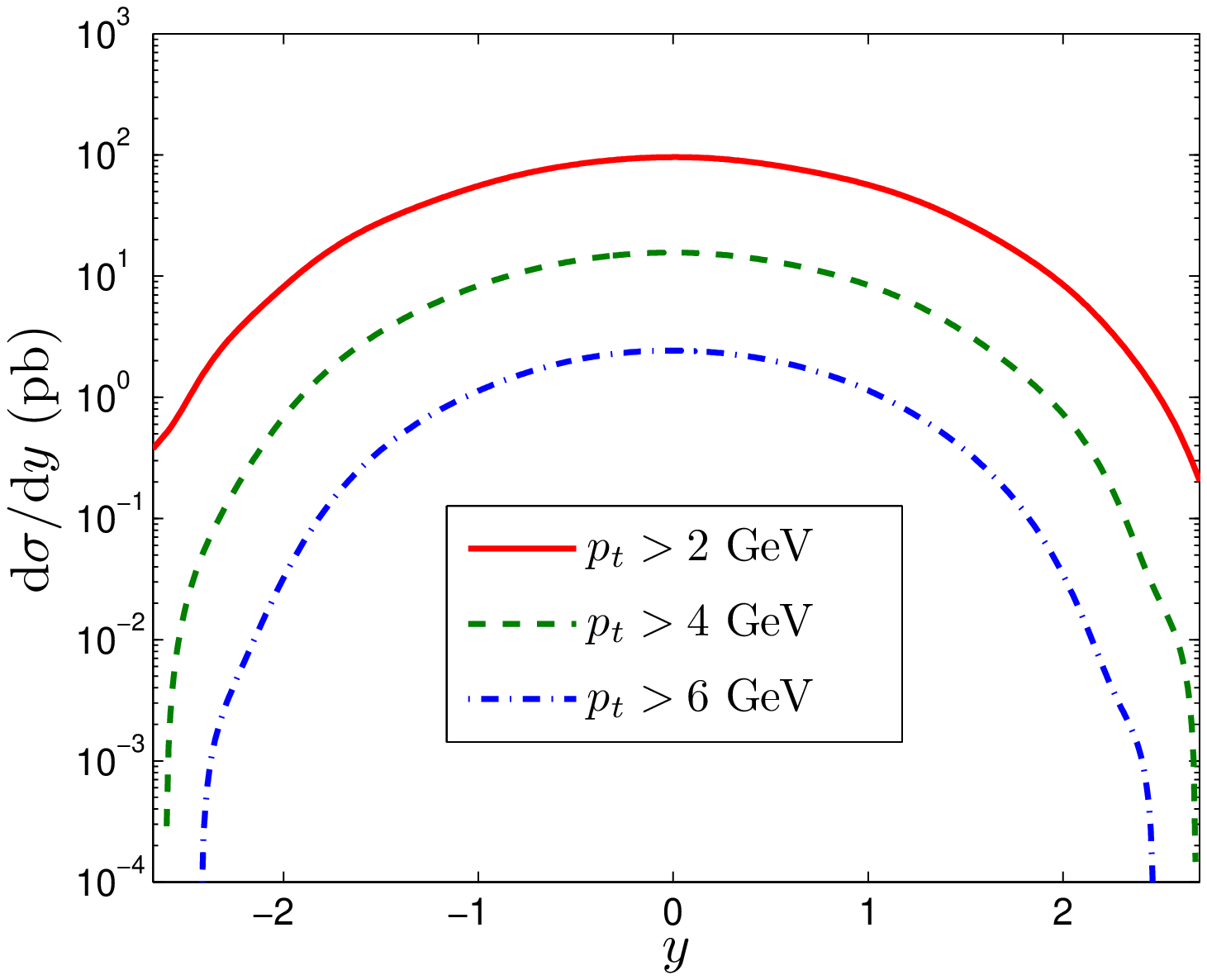}
\caption{The rapidity distributions of $\Xi_{cc}$ production in the gluon-gluon fusion mechanism with various $p_t$ cuts at the After@LHC, where the contributions via the $(cc)$-diquark states $[^3S_1]_{\bar{\textbf{3}}}$ and $[^1S_0]_{\textbf{6}}$ have been summed up.} \label{cutcc1}
\end{figure}

\begin{table}
\begin{tabular}{|c|c|c|c|}
\hline
~~ ~~ & ~$p_t>2$ GeV~ & ~$p_t>4$ GeV~ & ~$p_t>6$ GeV~ \\
\hline
$\sigma_{g+g}$ (pb) & 266 & 41.9 & 5.95 \\
\hline
$\sigma_{g+c}$ (pb) & 824 & 52.6 & 4.23 \\
\hline
$\sigma_{c+c}$ (pb) & 1.04 & 1.04 & 0.87 \\
\hline
\end{tabular}
\caption{Total cross sections for the $\Xi_{cc}$ production at the After@LHC with different $p_t$ cuts, where contributions from intermediate diquark states have been summed up.}
\label{ptcut}
\end{table}

Fig.(\ref{ptcc}) indicates that the baryon cross sections depend heavily on the baryon $p_t$. Total cross sections shall first increase with the increment of $p_t$ in small $p_t$ region and then drop down logarithmically in large $p_t$ region. Experimentally, it is also possible to apply more $p_t$ cuts to the production channels. To show how the baryon production depends on $p_t$, we put the $\Xi_{cc}$ cross sections for various production channels and various $p_t$ cuts in Table \ref{ptcut}. Table \ref{ptcut} shows: for $g+g$ channel, $\sigma_{gg}|_{p_t>0.2{\rm GeV}}: \sigma_{gg}|_{p_t>2{\rm GeV}}: \sigma_{gg}|_{p_t>4{\rm GeV}}: \sigma_{gg}|_{p_t>6{\rm GeV}} = 106:45:7:1$; for $g+c$ channel, $\sigma_{gc}|_{p_t>0.2{\rm GeV}}: \sigma_{gc}|_{p_t>2{\rm GeV}}: \sigma_{gc}|_{p_t>4{\rm GeV}}: \sigma_{gc}|_{p_t>6{\rm GeV}} = 836:195:12:1$.

Moreover, by taking the $\Xi_{cc}$ production via the gluon-gluon fusion as an explicit example, we present the rapidity distributions under various $p_t$ cuts in Fig.(\ref{cutcc1}), where the contributions via the $(cc)$-diquark states $[^3S_1]_{\bar{\textbf{3}}}$ and $[^1S_0]_{\textbf{6}}$ have been summed up.

\begin{figure}[htb]
\includegraphics[width=0.45\textwidth]{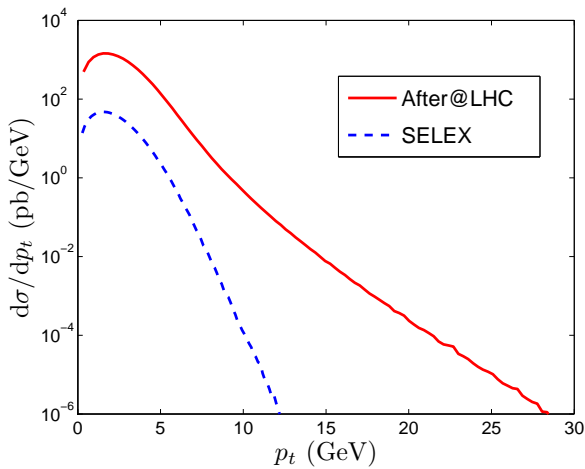}
\caption{A comparison of $\Xi_{cc}$ $p_t$ distributions at the After@LHC and the SELEX experiments, where all production channels have been summed up, $p_t>0.2$ GeV and no rapidity cut has been applied. }
\label{A&S}
\end{figure}

At the After@LHC, similar to the case of SELEX~\cite{genxicc}, the small $p_t$ behavior for the extrinsic charm mechanism becomes very important. Thus, the extrinsic charm mechanism should be taken into consideration for a sound estimation. As a comparison, we draw the $\Xi_{cc}$ $p_t$ distributions for both the SELEX and the After@LHC experiments in Fig.(\ref{A&S}), where all production channels have been summed up. In small $p_t$ region, the $\Xi_{cc}$ $p_t$ distribution at the After@LHC is larger than that of SELEX by about an order of magnitude. In large $p_t$ region, the $\Xi_{cc}$ $p_t$ distributions at both the After@LHC and the SELEX drop down logarithmically. Moreover, the $\Xi_{cc}$ $p_t$ distribution drops down more slowly at the After@LHC. This is due to a larger collision energy at the After@LHC: I) A larger collision energy means more small $x$ events, or equivalently more small $p_t$ events, can be generated at the After@LHC; II) It is noted that the PDFs for the incident partons drop down logarithmically with the increment of the parton fraction $x$, as can be explicitly shown by CT10~\cite{ct10}. Thus, more small $x$ events leads to more larger cross sections.

It is noted that the extrinsic charm mechanism via the channels $g+c\to \Xi_{cc}+\bar{c}$ and $c+c\to \Xi_{cc}+g$ provide dominant contribution to small $p_t$ events. To estimate the extrinsic charm mechanism for $\Xi_{cc}$ production, we define a ratio,
\begin{displaymath}
R = \frac{\sigma_{\rm tot}}{\sigma_{gg \to \Xi_{cc}(cc)_{ \bar{\textbf{3}}}[^3S_1]}},
\end{displaymath}
where $\sigma_{\rm tot}$ stands for the total cross section for all the concerned mechanisms in the $\Xi_{cc}$ production, and $\sigma_{gg \to \Xi_{cc}(cc)_{\bar{\textbf{3}}}[^3S_1]}$ is the total cross section for the channel $gg \to \Xi_{cc}(cc)_{\bar{\textbf{3}}}[^3S_1]+\bar{c}+\bar{c}$ only. At the SELEX, the ratio $R\simeq29$~\cite{genxicc}, which changes to $\simeq8$ at the After@LHC. Thus, the After@LHC is also possible to test the charmonium components in hadron.

There are some minor points for the present analysis:
\begin{itemize}
\item In the above estimations, we have fixed the renormalization scale $\mu_R$ to be the baryon's mass $M_{\Xi_{QQ'}}$. For the present leading-order pQCD calculation, the scale uncertainty is large, e.g., for the $\Xi_{cc}$ production via the $g+c$ channel, the scale uncertainty is $\pm22\%$ for $\mu_{R}\in[M_{\Xi_{cc}},\frac{\sqrt{s}}{2}]$, where $\sqrt{s}$ stands for the collision energy of the incident partons. However, by adopting an improved scale analysis suggested in Ref.\cite{pmc1}, which is based on the principle of maximum conformality (PMC)~\cite{pmc2,pmc3,pmc4,pmc5}, we shall obtain a smaller and hence a more reliable scale uncertainty $\pm 6\%$ for $\mu_{R}\in[M_{\Xi_{cc}},\frac{\sqrt{s}}{2}]$.

\item In the above discussions, we have not distinguished the light components in the baryon. More subtly, for $\Xi_{cc}$ production, after forming a $(cc)$-diquark, it will grab a light anti-quark (with soft gluons when necessary) from the hadron to form the final colorless doubly heavy baryon. According to the string model~\cite{pythia}, the possibility for grabbing the light (sea) quark from the hadron is $d :u :s \simeq 1:1:0.3$. If a $(cc)$-diquark is produced, it will fragment into $\Xi_{cc}^{+}$ with $43\%$ probability, $\Xi_{cc}^{++}$ with $43\%$ probability and $\Omega_{cc}^{+}$ with $14\%$ probability. Then, if enough $\Xi_{cc}$ events can be accumulated at the After@LHC, one may have chances to study the $\Xi_{cc}^{+,++}$ or $\Omega_{cc}^{+}$ separately from their decay products.

\item As has been estimated, there are two dominant $\Xi^+_{cc}$ decay channels, $\Xi^+_{cc} \to \Lambda^+_c ~\emph{K}^- ~\pi^+$ and $\Xi^+_{cc} \to \emph{p}~ \emph{D}^+ ~\emph{K}^-$. Setting $\Gamma_{1,2}$ to be the decay widths of these two channels, we have~\cite{selex1,selex2}: $\Gamma_1/\Gamma_2 = 0.36 \pm0.21$. As a rough estimation, if taking $\Gamma_1/\Gamma_2 = 0.36$ and the integrated luminosity to be 2 fb$^{-1}$, we shall have $1.0\times 10^{6}$ $\Xi^+_{cc}$ events from the first decay channel and $2.6\times 10^{6}$ $\Xi^+_{cc}$ events from the second decay channel at the After@LHC.

\end{itemize}

\section{Summary}

We have presented a detailed discussion on the doubly heavy baryon production at the suggested fixed target experiment After@LHC. For a fixed target experiment, more smaller $p_t$ events can be measured in comparison to the hadronic experiments as LHC and Tevatron. Since the baryon $p_t$ distributions drops down logarithmically in large $p_t$ region, this indicates that more baryon events can be produced at the After@LHC. If the integrated luminosity at the After@LHC reaches up to $2~{\rm fb}^{-1}$ per year, sizable $\Xi_{cc}$ and $\Xi_{bc}$ events can be generated, i.e., about $8.3\times10^6$/year $\Xi_{cc}$ and $1.8\times10^4$/year $\Xi_{bc}$ maybe observed for a small $p_t$ cut, $p_t>0.2$ GeV. If its luminosity can be improved further, we shall have much more baryon events available. Moreover, because of a larger collision energy than that of SELEX, more small $x$ events and more small $p_t$ events can be generated at the After@LHC. Thus, the After@LHC experiment shall provide a good platform to study the baryon properties and may greatly help to clarify the present SELEX puzzle.

\hspace{2cm}

{\bf Acknowledgement:} The authors would like to thank Stanley J. Brodsky and Jean-Philippe Lansberg for helpful suggestions and discussions. This work was supported in part by Natural Science Foundation of China under Grant No.11275280 and No.11347024, by the Program for New Century Excellent Talents in University under Grant No.NCET-10-0882, and by the Fundamental Research Funds for the Central Universities under Grant No.CQDXWL-2012-Z002.


\end{document}